\documentstyle [12pt]{article}
\def\be{\begin{equation}}
\def\ee{\end{equation}}
\def\bea{\begin{eqnarray}}
\def\eea{\end{eqnarray}}

\begin{document}

\begin{titlepage}

\title{Discrete and Backlund (!) transformations of SDYM system}


\author{A.N. Leznov\\Universidad Autonoma del Estado de Morelos,\\
CIICAp,Cuernavaca, Mexico}

\maketitle

\begin{abstract}

Symmetry of SDYM equations for "physically restricted solution" with hermitian group element $G=G^H$ in representation of Yang is described. Such transformation $D^B$ pass some PRS to the new one of the same kind. Transformation contain $2N_G$ arbirtrary functions of two independent arguments, where $N_G$ dimension of the gauge semi-simple algebra. These functions may be considered as additional parameters typical for Backlund aproach and by this reason we use term Backlund for this transformation. 

\end{abstract}
\end{titlepage}

\section{Introduction}

In a series of papers by the author starting in 90's  of the  previous century \cite{1},\cite{2},\cite{10} all nonlinear symmetries of integrable systems were united by the name of Backlund. In the present paper we would like to improve this misundestanding and
construct really the Backlund transformation for self-dual system of Yang-Mills.

The systems of these equations may be written in two equivalent form ("right" or "left"):
\be
G_{\bar y}G^{-1}=f_z, \quad  G_{\bar z}G^{-1}=-f_y \label{GFL}
\ee
\be
G^{-1}G_y=\bar f_{\bar z}, \quad  G^{-1}G_{\bar z}=-\bar f_{\bar y} \label{GFR}
\ee
where $G$ is the element of the gauge group and functions $f,\bar f$ takes values in corresponding semi-simple algebra (in general $\bar f$ not equal to $f^H$). 
From (\ref{GFL}) and (\ref{GFR}) it follows equations of the second order for group element $G$ ( representaion of Yang \cite{Y}) and algebra valued functions $f,\bar f$: 
\be
(G_{\bar y}G^{-1})_y+(G_{\bar z}G^{-1})_z=0,\quad f_{y,\bar y}+f_{z,\bar z}=
[f_z,f_y]\label{R}
\ee
\be
(G^{-1}G_y)_{\bar y}+(G^{-1}G_z)_{\bar z}=0,\quad \bar f_{y,\bar y}+\bar f_{z,\bar z}=
[\bar f_{\bar y},\bar f_{\bar z}]\label{L}
\ee
We call system (\ref{GFL}) or (\ref{GFR}) by the term "enlarged SDYM system" keeping in mind that in the case interesting for physical applications (for instance multi-instanton problem) it is necessary satisfy additional condition for its solution - namely the hermitianity of element $G=G^H$ and as a consequence $\bar f=f^H$. Such solution we call as "physically restricted solution" - PRS. This solution is connected with the obvious inner authomorphism of enlarged system $G\to G^H, f\to \bar f^H,\bar f\to f^H$. We call such authomorphism by term $\sigma_H$
(compare with \cite{AM}). The enlarged system is obviously invariant with the following
transformation $G\to \bar A(\bar y,\bar z)GA(y,z)$. Such symmetry we call the gauge one $D^G$. 
This symmetry conserves in the PRS case under additional condition $\bar A=A^H$.

Except of (\ref{R}) and (\ref{L}) from (\ref{GFL}) and (\ref{GFR}) it follows additionally
$$
f_{z,y}G=G\bar f_{\bar z,\bar y},\quad f_{z,z}G=-G\bar f_{\bar y,\bar y},\quad f_{y,y}G=-G\bar f_{\bar z,\bar z}
$$

\section{Discrete transformation of self-dual Yang-Mills system}

In this section we repeat the results of \cite{2}.
 
Keeping in mind (\ref{R}) it is not difficult to show that the following
system of equations for a group-valued element $S$ is self-consitent
(equalities of Maurer-Cartan are satisfied)
\be
S^{-1}S_y={1\over f_-}[X^+_M,f_y] -({1\over f_-})_{\bar z}X^+_M,\quad
S^{-1}S_z={1\over f_-}[X^+_M,f_z] +({1\over f_-})_{\bar y}X^+_M \label{S}
\ee
where $X^+_M$ maximal positive root of the algebra, $f_-$ - coefficient
on $X^-_M$ in decomposition $f$ function on the root system of the algebra.

The algebra-valued function $F$, satisfying the selfconsistent conditions
(in the sense of equality of the second mixed derivatives)
\be
F_y=S f_y S^{-1}-S_{\bar z}S^{-1}, \quad F_z=S f_z S^{-1}+S_{\bar y}S^{-1}
\label{DT}
\ee
also satisfy the self-dual Yang-Mills equations in the form (\ref{R}) and
new group value function $\tilde G=SG$ is solution of the same system in
the Yang's form. Now let us consider the transformaion of $\bar f$ functions.
From its definition (\ref{GFR}) it follows
$$
\bar F_{\bar z}=(SG)^{-1}(SG)_y=G^{-1}S^{-1}S_yG+G^{-1}G_y=\bar f_{\bar z}-({G^{-1}X^{+M}G\over 
f_-})_{\bar z}
$$
and the same expession with respect to differetiation with respect to $\bar y$. From which conclude that 
\be
\bar F=\bar f-{G^{-1}X^{+M}G\over f_-}\label{BAR}
\ee
up to function depending on $(z,y)$ arguments only. Thus formulae (\ref{S}),(\ref{DT}) and (\ref{BAR}) solve the problem of construction of new solution of enlarged system SDYM equations.
We will call such transformation as a left discrete transformation $D^L$ keeping in mind that in its result group element $G$ is multiplied on additional element $S$ from the left.

Repeating word by word the procedure above starting from (\ref{GFR}) we come to the right discrete transrmation. Formally this procedure is equivalent to fullfiling the hermitian congiguation of all formulae above under condition $G=G^H,\bar f=f^H,(X^{+M})^H=X^-$. Such transformation denotes as $D^R$. 

\section{Backlund transformation}

Now we would like to find  not the symmetry of the enlarged system
(\ref{GFR}),(\ref{GFL}) but independently the symmetry of PRS. This symmetry transformation can contain additional numerical parameters and after each its application we increase the number of parameters in the solution. This is exactly  Backlund's original idea. However, one has to keep in mind that, applied to the general solution of the SDYM equations, this transformation cannot add any new independent parameters, but can only change the initial functions on which the
general solution depends. In what follows we begin consideration from the case of $A_1$ gauge algebra. After these calculations the passing to the general case of arbitrary gauge algebra will be obvious.

In this section we would like to show that left and right discrete transformation applicated (in each order) to PRS lead also to PRS. More deep reason why this fact takes place at this moment is unknown to the author.
 
\subsection{The case $A_1$ gauge algebra}

Arbitrary element of $GL(2,C)$ looks as (the Gauss-Iwazava decomposition)
$$
G=e^{\alpha X_+}e^{\tau h}e^{\beta X_-},\quad [X_+,X_-]=h,\quad [h,X_{\pm}]=\pm 2 X_{\pm}
$$
Condition $G=G^H$ equivalent to restriction $\alpha=\beta^*, \tau=\tau^*$.
\be
G'G^{-1}=(\alpha'-2\tau'\alpha-\alpha^2\beta'e^{-2\tau}) X_++(\tau'+\beta'\alpha e^{-2\tau})h+
\beta'e^{-2\tau}X_-\label{PAR1}
\ee
where $f'$ is differentiation with respect to some variable.
\be
G^{-1}G'=(\beta'-2\tau'\beta-\beta^2\alpha'e^{-2\tau}) X_-+(\tau'+\alpha'\beta e^{-2\tau})h+
\alpha'e^{-2\tau}X_+\label{PAR2}
\ee
After application of these formulae to (\ref{R},) we obtain in component form (we present only the components necessary for further calcules)
\be
f^-_z=\beta_{\bar y}e^{-2\tau},\quad f^-_y=-\beta_{\bar z}e^{-2\tau},\quad \tau_{\bar y}=f^0_z-\alpha f_z,\quad -\tau_{\bar z}=f^0_y-\alpha f^-_y\label{P}
\ee
Formulae (\ref{S}) in the case under consideration look as
\be
S^{-1}S_y=(\ln f_-)_yh-( 2{f^0_y\over f_-} -({1\over f_-})_{\bar z})X^+,\quad
S^{-1}S_z=(\ln f_-)_zh-( 2{f^0_y\over f_-} +({1\over f_-})_{\bar y})X^+\label{S}
\ee
From which follows that group element $S$ belongs to the group of upper triangular matrices and can be parametrised in the form $S=e^{\rho h}e^{AX_+}$. Using (\ref{PAR2}) we obtain the system of equation for definition parameters $A,\rho$ (in what follows $f_-\equiv f^-$ and so on)
\be
\rho=\ln f_-,\quad (Af_-^2)_y=-2f^0_yf_-+f^-_{\bar z},\quad (Af_-^2)_z=-2f^0_zf_--f^-_{\bar y}
\label{MED}
\ee
The last system of equations for definition of $A$ is selfconsistent due to SDYM (\ref{R}).

Finally for transformed group element $G^t$ we obtain
$$
G^t=SG=e^{\rho h}e^{AX_+}e^{\alpha X_+}e^{\tau h}e^{\beta X_-}=
e^{(\alpha+A)f_-^2X_+}e^{(\tau +\ln f_-)h}e^{\beta X_-}
$$ 
and for $\bar f^t$ in connection with (\ref{BAR}) we obtain
\be
\bar f_+^t=\bar f_+-{e^{-2\tau}\over f_-},\quad \bar f_0^t=\bar f_0-{e^{-2\tau}\beta\over f_-}
\label{U}
\ee
We see that in finally expression for $G^t$ parameteres $A,\alpha$ arise only in combination
$(A+\alpha)f_-^2$ and for further consideration it will be more usefull to rewrite system (\ref{MED}) in equivalent form (using (\ref{P})): 
\be
((A+\alpha)f_-^2)_y=\alpha_y f_-^2+2\tau_{\bar z}f_-+f^-_{\bar z},\quad ((A+\alpha)f_-^2)_z=\alpha_zf_-^2-2\tau_{\bar y}f_--f^-_{\bar y}\label{MID}
\ee 
Of course in all previous formulae and in what follows it is not necessary to forget that initial solution is $\sigma_H$ invariant $(\alpha=\beta^*,\tau=\tau^*, \bar f=f^H)$. 
 
Now let us applicate transformation $D^R$ to solution obtained above. As it was mentioned above for this goal it is necessary in all formulae for $D^L$ fullfil operation of hermitian conjugation. 
On this way for group element we have now $S=e^{BX_-}e^{\theta}h$.
(\ref{MED}) transformed into
\be
\theta=\ln \bar f_+,\quad (B\bar f_+^2)_{\bar y}=-2\bar f^0_{\bar y}\bar f_++f^+_z,
$$
$$
(B\bar f_+^2)_{\bar z}=-2\bar f^0_{\bar z}\bar f_+-\bar f^+_y\label{MEDI}
\ee
and of course in all last forrmulae it is necessary consider elements of $\bar f$ in connection with (\ref{U}). Namely
$$ 
\bar f_+^t=f^*_--{e^{-2\tau}\over f_-},\quad \bar f_0^t=f^*_0-{e^{-2\tau}\beta\over f_-}
$$
After two transformations $D^RD^L$ finally group element takes the form
$$
G_{LR}=e^{(\alpha+A)f_-^2X_+}e^{(\tau +\ln f_-+\ln \bar f_+^t)h}e^{(B+\beta)(\bar f_+^t)^2 X_-}=
$$
$$
e^{(\alpha+A)f_-^2X_+}e^{(\tau +\ln (f_-f_-^*-e^{-2\tau}))h}e^{(B+\beta)(\bar f_+^t)^2 X_-}
$$
Now our goale is to show that $(\alpha+A)f_-^2$ and $(B+\beta)(\bar f_+^t)^2$ are complex conjuguated and thus $G_{LR}=G_{LR}^H$. Let us write down and transform left hand side of equation for definition of $(B+\beta)(\bar f_+^t)^2$ (compare with (\ref{MID}))
$$
(\beta(f_-^*-{e^{-2\tau}\over f_-})^2)_{\bar y}-2(f_0^*-{e^{-2\tau}\beta\over f_-})_{\bar y}(f_-^*-{e^{-2\tau}\over f_-})+(f_-^*-{e^{-2\tau}\over f_-})_z=
$$
Taking into account (\ref{P}) we rewrite the last relation as
$$
f^-_z(f_-^*-{e^{-2\tau}\over f_-})^2)+2[(\beta f_-^*-f_0^*)_{\bar y}-f^-_ze^{2\tau}(f_-^*-{e^{-2\tau}\over f_-})](f_-^*-{e^{-2\tau}\over f_-})+(f_-^*-{e^{-2\tau}\over f_-})_z=
$$
Using once more (\ref{P}) after simple algebraic manipulations we obtain
$$
f^-_z(f_-^*-{e^{-2\tau}\over f_-})^2)+2(\tau_z+{f^-_z\over f_-})(f_-^*-{e^{-2\tau}\over f_-})+(f_-^*-{e^{-2\tau}\over f_-})_z=
$$
$$
\beta_{\bar y} (f_-^*)^2+2\tau_zf_-^*+(f_-^*)_z
$$ 
which exactly coinsides with cmplex conjugation of right hand side of (\ref{MID}). Thus 
if introduce notation $D^B$ for Backlund transformation, then from the results of this section it follows
$$
D^B=D^LD^R=D^RD^L
$$
In validity of the fact of the mutual commutativity of the lef and right transformations it is not difficult verify by direct obvious calculations. And this fact follows directly from finally expresion for group element $G_{LR}$, which is symmetrical with respect to permutations of the parameters of $D^L$ and $D^R$ transformation.

\subsection{The case of arbitrary semisimple gauge algebra}

The discrete transformation $D^L$ in the case of arbitrary semisimple gauge algebra was introduced in \cite{2}. In what follows we partially repeat material of \cite{MP}.

We use the grading of the maximal root of the semisimple algebra. This means that all generators of the algebra may be distributed on the subspaces
with $\pm 2,\pm 1$ and $0$ graded indexes. Thus arbitrary element of the
semisimple algebra may be presented in the form
$$
f=f^{(+2}+f^{(+1}+f^{(0}+f^{(-1}+f^{(-2}, \quad [h,f^{(m}]=m f^{(m}, \quad
h=[X^+_M,X^-_M]
$$
The subspaces $f^{(\pm 2}\equiv f^{\pm}X^{\pm}_M$ are one dimensional.
Subspaces with $\pm 1$ graded indexes may decaupled $f^{(\pm 1}=f^{(\pm}_++
f^{(\pm 1}_-$ in such a way that $[f^{(\pm}_+,f^{(\pm}_-]=cX^{\pm}_M$ and all
generators in subspaces with the same additional index $\pm$ mutually
commutative. In other words $f^{(\pm}_+$ is the linear combination of the generators of $\pm 1$
gradded subspaces with generators with the indexes more then $[{M\over 2}]$; $f^{(\pm 1}_-$ linear combinations generators with indexes less then $[{M\over 2}]$. Generator with the index ${M\over 2}$ is absent ( the algebra is semisimple).
Corresponding group element $G$ may be reprersented as generalized Gauss-Iwazava decompositon in the form
\be
G=e^{\alpha X_+}e^{\alpha^{+1}_+}e^{\alpha^{+1}_-}e^{\tau h}g_0e^{\beta^{-1}_-}e^{\beta^{-1}_+}e^{\beta X_-}\label{PAR3}
\ee
Condition $G=G^H$ lead to restiction
$$
\alpha^*=\beta,\quad (\alpha^{+1}_+)^H=\beta^{-1}_+,\quad (\alpha^{+1}_-)^H=\beta^{-1}_-,\quad
\tau=\tau^*,\quad g_0=g^H_0
$$
All generators of $0$-gradded subspace except of $h$ are commuting with $X^{\mp M}$ and thus in (\ref{S}) right-hand sides is decomposed on $h,X^{+M}\equiv X^+$ and generators with $+1$ graded indexes. By this reason it is possible to seek solution for $S$ in the form
\be
S^R=e^{\tau h} e^{\epsilon^{(+1}_+} e^{\epsilon^{(+1}_-} e^{A X^+_M}
\label{SS}
\ee
where (and in what follows) $X^{\pm}_M,h$ maximal (minimal) root of the algebra with
corresponding Cartan element, $\epsilon^{(+1}_{\pm}$ elements of the
commutative subalgebras belonging to subspases with the grading index $+1$ (see comments few lines above).

Substituting (\ref{SS}) in the first equation (\ref{S}) and equating terms
in subspaces with the same grading indexes $(0,+1,+2)$ we obtain (in
what follows always $f_-\equiv f^-$)
$$
\tau=\ln f_-,\quad \epsilon^{(+1}=\epsilon^{(+1}_++\epsilon^{(+1}_-={[X^+_M,
f^{(-1}]\over f_-}
$$
And equation for $A$ function
\be
A_y X^+_M+[(\epsilon^{(+1}_+)_y,\epsilon^{(+1}_-]+{f^-_y\over f^-}
[\epsilon^{(+1}_,\epsilon^{(+1}_-]+2{f^-_y\over f^-}A X^+_M=
{[X^+_M,f^0_y]\over f^-}-{f^-_{\bar z}\over (f^-)^2}X^+_M \label{I}
\ee

Equations for algebra-valued function $f$ obviously can be partially resolved
as (see details in \cite{L87})
\be
f_{\bar z}+{1\over 2}[f,f_y]=R_y,\quad -f_{\bar y}+{1\over 2}[f,f_z]=R_z
\label{INT}
\ee
Calculating $-2$ graded component of these equations
(and twice commuting with $X^+_M$) we obtain
\be
-f^-_{\bar z}X^+_M+{1\over 2}[\epsilon^{(+1},(f^-\epsilon^{(+1})_y]-
{f^-_y\over 2}[X^+_M,f^0] +{f^-\over 2}[X^+_M,f^0_y]=-R^-_y X^+_M \label{III}
\ee
After substitution this expression into (\ref{I}) we obtain finally
\be
A f_- X^+_M+{f_-\over 2}[\epsilon^{(+1}_+,\epsilon^{(+1}_-]+
-{1\over 2}[X^+_M,f^0]=-{R^-\over f_-}X^+_M \label{II}
\ee
After multiplication (\ref{SS}) on initial element $G^L=SG$ and after some obvious manipulations we obtain in a result
\be 
G^L=e^{(\alpha+A)f_-^2 X_++[\epsilon^{+1}_-,\alpha^{+1}_+]f_-^2}e^{(\alpha^{+1}_++\epsilon^{+1}_+)f_-}e^{(\alpha^{+1}_-+\epsilon^{+1}_-)f_-}e^{(\tau +\ln f_-)h
}g_0e^{\beta^{-1}_-}e^{\beta^{-1}_+}e^{\beta X_-}\label{PARR}
\ee
Now if we want aplicate $D^R$ transformation to the obtained solution, at first it is necessary calculate $\bar f$ after $D^L$ transformation. With the help of (\ref{BAR}) for necessary for further calculations graded components of $\bar F$ we obtain
$$
\bar F^+=\bar f^+-{e^{-2\tau}\over f_-},\quad \bar F^{+1}=\bar f^{+1}-{[X^+,\beta^{-1}]\over f_-},
$$
$$
\bar F^0=\bar f^0-{\beta h+{1\over 2}([[X^+,\beta^{-1}]\beta^{-1}]]-[X^+,[\beta^{-1}_+,\beta^{-1}_-]]\over f_-}
$$
In connection with the comment above all calculations with respect to right transformation may be obtained by the formal complex conjugation the same of the left one. Thus right group-valued multiplicator necessary represent in the form
\be 
S^L=e^{B X^-} e^{\tilde \epsilon^{(+1}_-} e^{\tilde \epsilon^{(+1}_+} e^{\theta h}
\label{SSS}
\ee
Resolving corresponding equations we obtain
$$
\theta=\ln \bar F^+=\ln (\bar f^+-{e^{-2\tau}\over f_-}),\quad \tilde \epsilon^{-1}={[\bar F^{+1},X^-]\over \bar F^+}
$$
After multiplication $G^L$ (\ref{PARR}) from the right on $S^R$ (\ref{SSS}) the arguments of corresponding group exponents will the following one
$$
\tau+\ln f_-+\ln \bar F_+=\tau+\ln (f_-f_-^*-e^{-2\tau}) 
$$
\be
(\tilde \epsilon^{-1}_{\pm}+\beta^{-1}_{\pm})\bar F_+\label{ARG}
\ee
$$
(\beta+B)\bar F_+^2 X_++[\beta^{-1}_+,\tilde \epsilon^{-1}_-]\bar F_+^2 
$$
Let us consider second argument. We have in a consequence
$$
(\tilde \epsilon^{-1}+\beta^{-1})\bar F_+=[\bar F^{+1},X^-]+\beta^{-1}\bar F_+=
[\bar f^{+1}-{[X^+,\beta^{-1}]\over f_-},X_-]+\beta^{-1}(\bar f_+-{e^{-2\tau}\over f_-})=
$$
$$
[\bar f^{+1},X_-]+\beta^{-1}f_-^*=([X_+,f^{-1}]+\alpha^{+1}f_-)^H=((\epsilon^{+1}+\alpha^{+1})f_-)^H
$$
We remain to the reader the pleasure to prove that the third argument in (\ref{ARG}) exactly Hermitian congugate to
$$
(\alpha+A)f_-^2 X_++[\epsilon^{+1}_-,\alpha^{+1}_+]f_-^2 
$$
from which it follows that element $G_{LR}=G_{RL}=G^H_{RL}$ is hermitian one. Thus transformation constructed in this section is the Backlund transformation with respect to PRS of SDYM equations.

\section{Transformation of instanton charge density under the Backlund transformation}

In inroduced above notations instanton charge density is proportional to
$$
q=Trace(f_{yy}f_{zz}-f_{yz}f_{zy})
$$
Thus after left discrete transformation (\ref{S}) and (\ref{DT}) we have
$$
q^L=Trace(F_{yy}F_{zz}-F_{yz}F_{zy})
$$
For second derivatives $F$ we obtain consequently from (\ref{DT}) and (\ref{S})
$$
S^{-1}F_{yy}S=f_{yy}+[S^{-1}S_y,f_y]-(S^{-1}S_y)_{\bar z}=
$$
$$
f_{yy}+{1\over f_-}[[X^+,f_y]f_y]-2({1\over f_-})_{\bar z}[X^+,f_y]-{1\over f_-}[X^+,f_{y\bar z}]+({1\over f_-})_{\bar z,\bar z}X^+
$$
$$
S^{-1}F_{yz}S=f_{yz}+[S^{-1}S_z,f_y]-(S^{-1}S_z)_{\bar z}=
$$
$$
f_{yz}+{1\over f_-}[[X^+,f_z]f_y]+({1\over f_-})_{\bar y}[X^+,f_y]-({1\over f_-})_{\bar z}[X^+,f_z]-{1\over f_-}[X^+,f_{z\bar z}]-{1\over f_-})_{\bar y,\bar z}X^+
$$
$$
S^{-1}F_{zy}S=f_{zy}+[S^{-1}S_y,f_z]+(S^{-1}S_y)_{\bar y}=
$$
$$
f_{yz}+{1\over f_-}[[X^+,f_y]f_z]-({1\over f_-})_{\bar z}[X^+,f_z]+({1\over f_-})_{\bar y}[X^+,f_y]-{1\over f_-}[X^+,f_{y\bar y}]-{1\over f_-})_{\bar y,\bar z}X^+
$$
$$
S^{-1}F_{zz}S=f_{zz}+[S^{-1}S_z,f_z]+(S^{-1}S_z)_{\bar y}=
$$
$$
f_{zz}+{1\over f_-}[[X^+,f_z]f_z]+2({1\over f_-})_{\bar y}[X^+,f_z]+{1\over f_-}[X^+,f_{z\bar y}]+({1\over f_-})_{\bar y,\bar y}X^+
$$
After not combersome calculations (without mistakes) we obtain
$$
q^L=q^{in}+(\partial^2_{y,\bar y}+\partial^2_{z,\bar z})(\partial^2_{y,\bar y}+\partial^2_{z,\bar z})\ln f_-
$$
Summating this result with obtained in the result of $D^R$ transformation we have 
\be
q^B=q^{in}+(\partial^2_{y,\bar y}+\partial^2_{z,\bar z})(\partial^2_{y,\bar y}+\partial^2_{z,\bar z})\ln (f_-f_-^*-e^{-2\tau})\label{IC}
\ee
 
\section{Parameters of Backlund transformation. 1-instanton solution}

On the first look constucted above $D^B$ transfortmation does't contain any additional parameters. But this is not so. Let we have some PRS $G=G^H$. Now this solution may be transformed by the gauge one $G\to \Psi G \Psi^H$ and after this applicate to such solution Backlund transformation. The number of additonal parameters arises on this way exactly equal to $2N_G$ arbitrary functions of two variables, where $N_G$ dimension of the gauge algebra (group).

To clarify situation let us consider zero instanton solution for $A_1$ gauge algebra. Such solution is obvious $G=1$. Now let us fullfile gauge transformation. 
\be
G=\bar \Psi(\bar y,\bar z)\Psi(y,z)\label{GEN}
\ee
Using (\ref{GFL}) we obtain correspoding $f$ ($\bar \Psi_{21}=\bar \theta,\bar \Psi_{22}=\bar \phi)$)
$$
f_-=\bar \phi(z\partial_{\bar y}-y\partial_{\bar z})\bar \theta - \bar \theta(z\partial_{\bar y}-y\partial_{\bar z})\bar \phi+\bar \psi (\bar y,\bar z)
$$
For solution (\ref{GEN})
$$
e^{-\tau}=G_{2,2}=\theta\bar \theta+\phi\bar \phi
$$
Let us choose $\bar \theta=\bar y,\bar \phi=\bar z,\bar \psi=a=constant$. This leeds to
$$
f_-=y\bar y+z\bar z+a,\quad e^{-\tau}=y\bar y+z\bar z
$$
Instantone density for the initial solution (\ref{GEN}) equal to zero and thus for instanton density after Backlund transformation (\ref{IC}) we obtain
$$
q^B=(\partial^2_{y,\bar y}+\partial^2_{z,\bar z})(\partial^2_{y,\bar y}+\partial^2_{z,\bar z})\ln (a\bar a+(a+\bar a)(y\bar y+z\bar z))
$$
this is exactly wellknown one instanton charge density of the ussual theory.

\section{Outlook}

To the best of our knowledges the symmetry of PRS of SDYM system of the present paper was not marked in the literature up to now. Of course the most interesting is the question will it be possible with the help of this symmetry to construct multi-instanton solution in the form different from famous ADHM anzats \cite{ADHM}.

Author would like to do some comments on this subject. For solution of the multi-instanton problem it is sufficient to investigate only two instanton configuration by the methods of the previous section. One instanton solution is known up to 3-functions of two independent variables (in construction above was used only matrix elements elements $G_{21},G_{22}$ of the group element $G$ and element $\bar \Psi_{21}$ of $A_1$ algebra-valued function $\Psi$). The new Backlund transformation add to this 6 functions of two independent arguments. All corresponding formulae are explicit. Question arises: if it is possible by the sucsessefull choise of these
functions come to instanton charge equal to 2? If after fulfiling such prosedure the answer would be positive and the number of arbitrary parameters will be equal to 8, then each Backlund transformation will be numerated by 8 numerical parameters $J$. In ussual case
Backlund transformation are always commutative \cite{AM}. This means that result of consequent application of n Backlund transformation  $ D^B_{J_1}D^B_{J_2} ....D^B_{J_n}$ to vaccum solution is symmetrical with respect to mutual permuation of the set of the indexes $J_i\to J_j$. On this way the n instanton solution would be constructed.

Author shure that in the nearest time this problem will be solved but do not know in positive or negative sence.

\end{document}